\documentstyle [12pt,axodraw] {article}

\catcode`@=12
\begin{document}
\thispagestyle{empty}
\begin{flushright} UCRHEP-T245\\May 1999\
\end{flushright}
\vspace{0.5in}
\begin{center}
{\Large \bf Simple Connection Between Atmospheric\\
and Solar Neutrino Vacuum Oscillations\\}
\vspace{1.3in}
{\bf Ernest Ma\\}
\vspace{0.3in}
{\sl Department of Physics\\}
{\sl University of California\\}
{\sl Riverside, California 92521\\}
\vspace{1.3in}
\end{center}
\begin{abstract}\ 
Extending the minimal standard model of particle interactions (without 
right-handed singlet neutrinos) to include a heavy scalar triplet $\xi$ to 
obtain nonzero Majorana neutrino masses, I derive the following simple 
realistic connection between atmospheric and solar neutrino vacuum 
oscillations: $(\Delta m^2)_{sol} (\Delta m^2)_{atm} / m_\nu^4 ~(\sin^2 2 
\theta)_{atm} = 2 I^2$, where $m_\nu$ is the assumed common approximate 
mass of each neutrino (which may be suitable for hot dark matter) and $I = 
(3/16 \pi^2)(G_F/\sqrt 2) m_\tau^2 \ln (m_\xi^2/m_W^2)$ comes from the 
radiative splitting of the degeneracy due to the charged leptons.
\end{abstract}
\newpage
\baselineskip 24pt

There is now a vast literature on models of neutrino oscillations \cite{1}. 
Most try to understand why atmospheric neutrino oscillations \cite{2} of 
$\nu_\mu (\bar \nu_\mu)$ to $\nu_\tau (\bar \nu_\tau)$ require near-maximal 
mixing \cite{3}.  Many also suggest that solar neutrino oscillations \cite{4} 
of $\nu_e$ to a linear combination of $\nu_\mu$ and $\nu_\tau$ should have 
near-maximal mixing as well \cite{5}.  Both are possible in the context of 
three nearly mass-degenerate neutrinos \cite{6,7} which could then be 
considered as candidates for hot dark matter \cite{8}.

Recently it has been pointed out \cite{9} that if all three neutrinos obtain 
equal Majorana masses of order 1 eV from the canonical seesaw mechanism 
\cite{10}, then their splitting due to the different charged-lepton masses 
from the two-loop exchange of two $W$ bosons \cite{11} is of the right 
magnitude for solar neutrino vacuum oscillations.  However, the inclusion 
of atmospheric neutrino oscillations has to be rather {\it ad hoc} in this 
case.  In fact, it is rare indeed that any {\it bona fide} model of neutrino 
masses even gets a relationship between the mass difference of one 
oscillation and that of another.  [One exception is the recently proposed 
model \cite{12} of radiative masses for $\nu_e, \nu_\mu, \nu_\tau$, plus a 
singlet (sterile) neutrino $\nu_s$, which explains atmospheric and solar 
neutrino oscillations as well as the $\bar \nu_\mu (\nu_\mu)$ to $\bar \nu_e 
(\nu_e)$ data of the LSND (Liquid Scintillator Neutrino Detector) experiment 
\cite{13}.  It has the successful relationship $(\Delta m^2)_{atm} \simeq 
2 [(\Delta m^2)_{sol} (\Delta m^2)_{LSND}]^{1/2}$, where $(\Delta m^2)_{sol}$ 
refers to the matter-enhanced solution \cite{14} of the solar neutrino 
deficit.]

In this note I will present the most economical model to date of neutrino 
masses which has the following simple realistic connection between atmospheric 
and solar neutrino vacuum oscillations:
\begin{equation}
{(\Delta m^2)_{sol} (\Delta m^2)_{atm} \over m_\nu^4 ~(\sin^2 2 \theta)_{atm}} 
= 2 I^2 = 4.9 \times 10^{-13} \left( \ln {m_\xi^2 \over m_W^2} \right)^2,
\end{equation}
where $m_\nu$ is the assumed common approximate mass of each neutrino, 
$m_\xi$ is the mass of a heavy scalar triplet, and
\begin{equation}
I = {3 G_F m_\tau^2 \over 16 \pi^2 \sqrt 2} ~\ln {m_\xi^2 \over m_W^2} 
\end{equation}
comes from the one-loop radiative splitting of the degeneracy due to the 
charged leptons, as explained below.  Numerically, let $m_\nu = 0.6$ eV, 
$(\sin^2 2 \theta)_{atm} = 1$, and $m_\xi = 1$ TeV, then Eq.~(1) is satisfied 
with the best fit values of $(\Delta m^2)_{sol} = 4.0 \times 
10^{-10}$ eV$^2$ and $(\Delta m^2)_{atm} = 4.0 \times 10^{-3}$ eV$^2$.

To start with, the minimal standard model (without right-handed singlet 
neutrinos) is extended to include a heavy scalar triplet $\xi = (\xi^{++}, 
\xi^+, \xi^0)$, where $m_\xi^2 >> m_W^2$ is assumed.  This provides the three 
neutrinos $\nu_e, \nu_\mu, \nu_\tau$ with small Majorana masses \cite{15}. 
As emphasized recently \cite{16}, such an alternative is as simple and natural 
as the canonical seesaw mechanism \cite{10} which was used in Ref.~[9].  Now 
let there be a discrete $S_3$ symmetry (which has irreducible representations 
\underline {2}, \underline {1}, and \underline {1$'$}) such that $\xi$ is a 
\underline {1} and the standard Higgs doublet $\Phi = (\phi^+, \phi^0)$ is 
also a \underline {1}, whereas two of the lepton doublets form a \underline 
{2} and the third is a \underline {1} or \underline {1$'$}.  The relevant 
terms in the interaction 
Lagrangian are then given by
\begin{equation}
{\cal L}_{int} = \xi^0 [f_0 (\nu_1 \nu_2 + \nu_2 \nu_1) + f_3 \nu_3 \nu_3] 
+ \mu \bar \xi^0 \phi^0 \phi^0 + h.c.
\end{equation}
The field $\xi^0$ acquires a naturally small vacuum expectation value 
\cite{15} $u \simeq - \mu \langle \phi^0 \rangle^2 / m_\xi^2$ and the 
$3 \times 3$ Majorana neutrino mass matrix is of the form
\begin{equation}
{\cal M}_\nu = \left( \begin{array}{c@{\quad}c@{\quad}c} 0 & m_0 & 0 \\ 
m_0 & 0 & 0 \\ 0 & 0 & m_3 \end{array} \right),
\end{equation}
where $m_0 = 2 f_0 u$ and $m_3 = 2 f_3 u$.  Actually, the difference between 
$m_0$ and $m_3$ will be assumed small compared to either $m_0$ or $m_3$ in 
the following, {\it i.e.} each neutrino is accorded an approximate common mass 
$m_\nu$.

The neutrinos are now identified with their charged-lepton partners as 
follows:
\begin{equation}
\nu_1 = \nu_e, ~~~ \nu_2 = c \nu_\mu - s \nu_\tau, ~~~ \nu_3 = c \nu_\tau + 
s \nu_\mu,
\end{equation}
where $s \equiv \sin \theta$ and $c \equiv \cos \theta$.  This construction 
is made to accommodate the atmospheric data \cite{2} as $\nu_\mu - 
\nu_\tau$ oscillations with $\sin^2 2 \theta = 4 s^2 c^2$ and $\Delta m^2 = 
m_0^2 - m_3^2$.  At this point, the eigenvalues of ${\cal M}_\nu$ of Eq.~(4) 
are $-m_0, m_0$, and $m_3$.  However, since the charged-lepton masses break 
the assumed $S_3$ symmetry, the two-fold degeneracy of the $\nu_1 - \nu_2$ 
sector is broken radiatively in one loop.  There are two effects.  One is a 
finite correction to the mass matrix, as shown in Figure 1.  The other is a 
renormalization of the coupling matrix \cite{17} from the shift in mass scale 
from $m_\xi$ to $m_W$.  As expected, the dominant contribution comes from the 
$\tau$ Yukawa coupling.  The two contributions are naturally of the same 
texture and are easily calculated to be $4I/3$ and $-I/3$ respectively, where 
$I$ is already given by Eq.~(2).  The mass matrix ${\cal M}_\nu$ is now 
corrected to read
\begin{equation}
{\cal M}_\nu = \left( \begin{array}{c@{\quad}c@{\quad}c} 0 & m_0 (1+s^2I) & 
-scm_0 I \\ m_0 (1+s^2I) & 0 & -scm_3 I \\ -scm_0 I & -scm_3 I & m_3 (1+2c^2I) 
\end{array} \right).
\end{equation}
The two-fold degeneracy of the $\nu_1 - \nu_2$ sector is then lifted, with 
the following mass eigenvalues:
\begin{equation}
-m_0 (1+s^2I) - {s^2 c^2 (m_0-m_3)^2 I^2 \over 2 (m_0+m_3)}, ~~~ 
m_0 (1+s^2I) + {s^2 c^2 (m_0+m_3)^2 I^2 \over 2 (m_0-m_3)},
\end{equation}
where $I^2 << (m_0 - m_3)^2/(m_0 + m_3)^2$ has been used, being justified 
numerically. Hence their mass-squared difference is
\begin{equation}
\Delta m^2 \simeq s^2 c^2 m_0 I^2 \left[ {(m_0+m_3)^2 \over m_0-m_3} - 
{(m_0-m_3)^2 \over m_0+m_3} \right] \simeq {8 s^2 c^2 I^2 m_\nu^4 \over m_0^2 
-m_3^2},
\end{equation}
where $m_\nu \simeq m_0 \simeq m_3$ has been used.  Identifying this with 
solar neutrino vacuum oscillations then yields Eq.~(1).

In the above, the choice $\nu_1 = \nu_e$ leads to $(\sin^2 2 \theta)_{sol} 
= 1$.  The eigenstates of ${\cal M}_\nu$ from Eq.~(4) or Eq.~(6) are the 
same to first order:
\begin{equation}
{1 \over \sqrt 2} (\nu_e - c \nu_\mu + s \nu_\tau), ~~~ {1 \over \sqrt 2} 
(\nu_e + c \nu_\mu - s \nu_\tau), ~~~ s \nu_\mu + c \nu_\tau.
\end{equation}
For $s = c = 1/\sqrt 2$, the so-called bimaximal mixing solution \cite{5} 
of neutrino oscillations is obtained.  With the assumed form of Eq.~(4), 
it is also worth noting that renormalization effects due to the $\tau$ and 
$\mu$ Yukawa couplings do not affect the degeneracy of the $\nu_1 - \nu_2$ 
sector to first order.  This is why $(\Delta m^2)_{sol}$ can be small enough 
here to be suitable for vacuum oscillations. The zero $\nu_e - \nu_e$ entry in 
the neutrino mass matrix is crucial for the validity of Eq.~(7) and has been 
chosen to avoid neutrinoless double beta decay \cite{18}.  This is an 
important constraint as long as $m_\nu$ is greater than about 1 eV, which 
used to be a desirable feature as a component of dark matter \cite{8}.  
However, with the recent observation of a nonzero cosmological constant 
\cite{19}, whereas $m_\nu$ is probably still needed for large-scale structure 
formation in the universe, its magnitude can be much smaller.  In general, 
$\nu_1$ may be a linear combination of $\nu_e$, $\nu_\mu$, and $\nu_\tau$, 
but it has to be predominantly $\nu_e$.  Otherwise, $m_\tau$ (and $m_\mu$) 
radiative contributions would appear in the diagonal entries of Eq.~(6) and 
modify Eqs.~(7) and (8). 
For illustration, the values $m_\nu = 0.6$ eV and $m_\xi = 1$ TeV have been 
used.  It may be argued that $m_\xi$ is naturally of order $10^{13}$ GeV or 
greater \cite{15}, in which case $m_\nu$ should be somewhat smaller.  More 
precisely,
\begin{equation}
m_\nu \sim 1.3 ~{\rm eV} \left[ \ln {m_\xi^2 \over m_W^2} 
\right]^{-{1 \over 2}}.
\end{equation}
For $m_\xi = 10^{13}$ GeV, the required $m_\nu$ is then about 0.18 eV.

The charged-lepton mass matrix which accompanies ${\cal M}_\nu$ of Eq.~(4) 
is not uniquely defined, because only the left-handed fields are correlated 
with it.  Nevertheless, $S_3$ is clearly violated.  So far, I have not 
identified the origin of this violation.  It may simply be explicit, or 
it may be spontaneous, in the sense that it occurs only when the electroweak 
gauge symmetry is broken.  An example of the latter is the following model. 
Under $S_3$, let
\begin{equation}
\left[ \left( \begin{array} {c} \nu_1 \\ l_1 \end{array} \right)_L, \left( 
\begin{array} {c} \nu_2 \\ l_2 \end{array} \right)_L \right] \sim 2, ~~~ 
\left( \begin{array} {c} \nu_3 \\ l_3 \end{array} \right)_L \sim 1, ~~~ 
[ l^c_{1L}, l^c_{2L} ] \sim 2, ~~~ l^c_{3L} \sim 1,
\end{equation}
\begin{equation}
\left[ \left( \begin{array} {c} \phi_1^0 \\ \phi_1^- \end{array} \right), 
\left( \begin{array} {c} \phi_2^0 \\ \phi_2^- \end{array} \right) \right] 
\sim 2, ~~~ \left( \begin{array} {c} \phi_3^0 \\ \phi_3^- \end{array} \right) 
\sim 1, ~~~ (\xi^{++}, \xi^+, \xi^0) \sim 1.
\end{equation}
With $\langle \xi^0 \rangle \neq 0$, ${\cal M}_\nu$ of Eq.~(4) is obtained, 
whereas ${\cal M}_l$ is now given by
\begin{equation}
{\cal M}_l = \left[ \begin{array} {c@{\quad}c@{\quad}c} h_1 \langle \phi_1^0 
\rangle & h_2 \langle \phi_3^0 \rangle & h_3 \langle \phi_2^0 \rangle \\ 
h_2 \langle \phi_3^0 \rangle & h_1 \langle \phi_2^0 \rangle & h_3 \langle 
\phi_1^0 \rangle \\ h_4 \langle \phi_2^0 \rangle & h_4 \langle \phi_1^0 
\rangle & h_5 \langle \phi_3^0 \rangle \end{array} \right],
\end{equation}
where $h_{1,2,3,4,5}$ are the couplings of all possible Yukawa terms 
invariant under $S_3$.  Before electroweak symmetry breaking, charged-lepton 
masses as well as neutrino masses are zero, as in the standard model.  After 
electroweak symmetry breaking, let $\langle \phi_2^0 \rangle << \langle 
\phi_1^0 \rangle$, then $S_3$ is also broken in ${\cal M}_l$ at tree level 
but not in ${\cal M}_\nu$.  Radiative corrections then break $S_3$ in 
${\cal M}_\nu$ as shown in this paper.  In the limit $h_2 \to 0$ and 
$\langle \phi_2^0 \rangle \to 0$, $e_L$ is indeed separated from the 
$\mu_L - \tau_L$ sector and Eq.~(5) holds as desired.  In terms of fine 
tuning, this model is no worse than the standard model which also requires 
arbitrary Yukawa couplings to fix the charged-lepton masses.

In conclusion, I have presented in this note a new and economical extension 
of the minimal standard model, where a heavy scalar triplet $\xi$ is 
added to provide the three known neutrinos with nonzero Majorana masses. 
This replaces the usual method of adding three heavy right-handed neutrino 
singlets.  A discrete $S_3$ symmetry is then assumed so that 
two neutrinos are degenerate in mass, with their splitting controlled by 
one-loop radiative corrections.  This results in a simple realistic 
connection between atmospheric and solar neutrino vacuum oscillations as 
given by Eq.~(1).  It is consistent with the present data and will get tested 
further as more data become available in the near future from planned 
experiments in neutrino oscillations, neutrinoless double beta decay, 
neutrino mass, searches for dark matter and for new particles in high-energy 
accelerators.

\begin{center} {ACKNOWLEDGEMENT}
\end{center}

This work was supported in part by the U.~S.~Department of Energy under 
Grant No.~DE-FG03-94ER40837.

\newpage
\bibliographystyle {unsrt}

\newpage
\begin{center}
\begin{picture}(360,200)(0,0)
\ArrowLine(30,0)(90,0)
\Text(60,-8)[c]{$\nu_i$}
\ArrowLine(180,0)(90,0)
\Text(135,-8)[c]{$\tau_L$}
\ArrowLine(270,0)(180,0)
\Text(225,-8)[c]{$\tau_R$}
\ArrowLine(330,0)(270,0)
\Text(300,-8)[c]{$\nu_\tau$}
\DashArrowLine(180,-40)(180,0)6
\Text(180,-50)[c]{$\langle \phi^0 \rangle$}
\DashArrowLine(180,97)(180,57)6
\Text(180,106)[c]{$\langle \phi^0 \rangle$}
\DashArrowArcn(180,-45)(101,154,90)6
\Text(130,52)[c]{$\xi^-$}
\DashArrowArcn(180,-45)(101,90,26)6
\Text(240,52)[c]{$\phi^-$}
\end{picture}
\vskip 2.0in
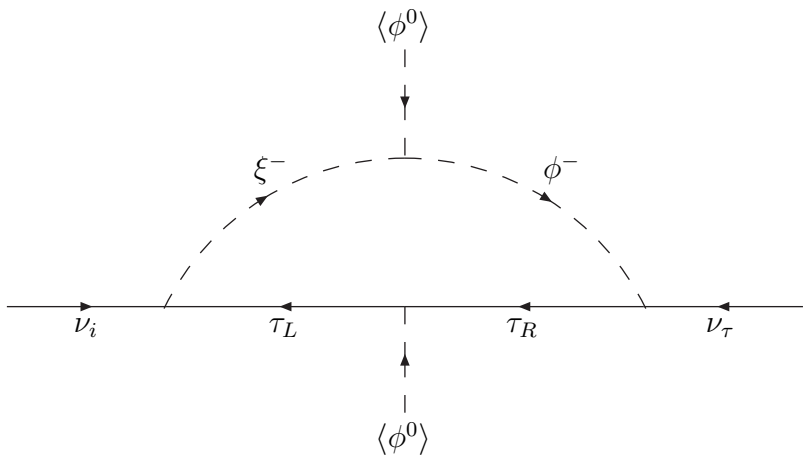
{\bf Fig.~1.} ~ One-loop radiative breaking of neutrino mass degeneracy.
\end{center}
\end{document}